\documentstyle[12pt]{article}

\def\be{\begin{equation}}
\def\ee{\end{equation}}
\def\bea{\begin{eqnarray}}
\def\eea{\end{eqnarray}}

\begin{document}

\centerline{\bf SYMMETRIES AND STAGGERING EFFECTS} 

\centerline{\bf IN NUCLEAR ROTATIONAL SPECTRA}

\bigskip
\centerline{NIKOLAY MINKOV, S. B. DRENSKA, P. P. RAYCHEV, R. P. ROUSSEV}

\centerline{Institute for Nuclear Research and Nuclear Energy}

\centerline{72 Tzarigrad Road, 1784 Sofia, Bulgaria}

\medskip
\centerline{DENNIS BONATSOS}

\centerline{Institute of Nuclear Physics, N.C.S.R. ``Demokritos''}

\centerline{GR-15310 Aghia Paraskevi, Attiki, Greece}

\bigskip

\centerline{\bf Abstract}

We study the fine structure of nuclear
rotational spectra on the basis of both dynamical and discrete
symmetry approaches. In this framework we show that the odd--even
($\Delta I=1$) staggering effects observed in various rotational
bands carry detailed information about the collective band-mixing
interactions and the collective shape properties of heavy nuclei.

\section{Introduction}

The odd--even staggering effect is known as a zigzagging behavior
of the nuclear inertial parameter between the odd and the even
angular momentum states of a rotational band. It due to a
relative displacement of the odd angular momentum levels  with
respect to the even ones \cite{BM75}. The analysis and the
interpretation of this effect are of current interest since it
carries detailed information about the fine properties of nuclear
collective dynamics in different regions of nuclear chart.

We study the odd--even staggering effects in nuclear spectra in
terms of an appropriately defined quantity
\begin{equation}
\mbox{Stg}(I)= 6\Delta E(I)-4\Delta E(I-1)-4\Delta E(I+1)+ \Delta
E(I+2)+\Delta E(I-2) \ , \label{stag}
\end{equation}
which is  the discrete approximation of the fourth derivative of
the function $\Delta E(I)=E(I+1)-E(I)$, i.e. the fifth derivative
of the rotational band energy $E(I)$. In the present paper it
will be shown that the above quantity is very sensitive to the
fine structure of rotational spectra and provides spectacular
$\Delta I=1$ staggering patterns (zigzagging behavior of the
function $Stg(I)$ with clearly defined zero reference) in various
rotational bands. On this basis we propose relevant theoretical
analysis of the $\Delta I=1$ staggering effects in collective
$\gamma$ rotational bands and nuclear octupole bands in reference
to the underlying symmetries of nuclear collective interactions.

\section{SU(3) dynamical symmetry and
$\Delta I =1$ staggering in heavy deformed nuclei}

We have found \cite{MDRRB00} that Eq.~(\ref{stag}) provides well
developed staggering patterns in the $\gamma$ bands of the nuclei
$^{156}$Gd, $^{156,160,162}$Dy, $^{162-166}$Er, $^{170}$Yb and
$^{228,232}$Th. We demonstrated that the observed effect can be
interpreted as the result of the interaction of the $\gamma$ band
with the ground band in the framework of a Vector Boson Model
with SU(3) dynamical symmetry \cite{MDRRBSU3}. In this model the
two bands are coupled into the same SU(3) multiplet, which
provides the following $\gamma$- band energy expression
\cite{MDRRB00}:
\begin{eqnarray}
E^{\gamma}(I)&=&2B+AI(I+1)
+B[\sqrt{1+aI(I+1)+bI^{2}(I+1)^{2}} \nonumber \\
&-&\left. CI(I+1)-1\right] \left(\frac{1+(-1)^I}{2}\right) \ ,
\label{Egamma}
\end{eqnarray}
where the quantities $A$, $B$, $C$ , $a$ and $b$ are determined by
the effective model interaction.  Eq.~(\ref{Egamma}) reproduces
successfully the $\Delta I =1$ effect in all considered nuclei. In
Fig. 1(a) the experimental and the theoretical staggering patterns
obtained for the $\gamma$ band of $^{166}$Er are illustrated.

Our theoretical analysis of the staggering patterns observed in
rare earth nuclei provide detailed information about the fine
behavior of the ground--$\gamma$ band mixing interaction in
dependence on the nuclear shell structure in rotational regions.
In addition, it suggests a detailed test and relevant comparison
of the different kinds of dynamical symmetry schemes.

\section{Octahedron point symmetry and $\Delta I =1$ staggering
in octupole bands}

We propose a study of the fine structure of collective rotational
bands with a presence of octupole degrees of freedom through the
formalism of the octahedron $(O)$ point symmetry group. Based on
the irreducible representations (irreps) of this group
\cite{Ham91}, we have constructed a collective Hamiltonian of a
system with octupole correlations: $
\hat{H}_{oct}=\hat{H}_{A_{2}}+
\sum_{r=1}^{2}\sum_{i=1}^{3}\hat{H}_{F_{r}(i)}$ with
\begin{eqnarray}
\hat{H}_{A_{2}}&=&{a}_{2}\frac{1}{4}
[(\hat{I}_x\hat{I}_y+\hat{I}_y\hat{I}_x)\hat{I}_z+
\hat{I}_z(\hat{I}_x\hat{I}_y+\hat{I}_y\hat{I}_x)] \ ,
\label{HA} \\
\hat{H}_{F_{1}(1)}&=&\frac{1}{2}{f}_{11}
\hat{I}_z(5\hat{I}_z^{2}-3\hat{I}^{2}) \ ,
\label{HF11} \\
\hat{H}_{F_{1}(2)}&=&\frac{1}{2}{f}_{12}
(5\hat{I}_x^{3}-3\hat{I}_x\hat{I}^{2}) \ ,
\label{HF12} \\
\hat{H}_{F_{1}(3)}&=&\frac{1}{2}{f}_{13}
(5\hat{I}_y^{3}-3\hat{I}_y\hat{I}^{2}) \ ,
\label{HF13} \\
\hat{H}_{F_{2}(1)}&=&{f}_{21}\frac{1}{2}
[\hat{I}_z(\hat{I}_x^{2}-\hat{I}_y^{2})+
(\hat{I}_x^{2}-\hat{I}_y^{2})\hat{I}_z] \ ,
\label{HF21} \\
\hat{H}_{F_2(2)}&=&{f}_{22} (\hat{I}_x\hat{I}^{2}-\hat{I}_x^{3}-
\hat{I}_x\hat{I}_z^{2}-\hat{I}_z^{2}\hat{I}_x) \ ,
\label{HF22} \\
\hat{H}_{F_2(3)}&=&{f}_{23}
(\hat{I}_y\hat{I}_z^{2}+\hat{I}_z^{2}\hat{I}_y+
\hat{I}_y^{3}-\hat{I}_y\hat{I}^{2})\ . \label{HF23}
\end{eqnarray}
Here, the first term $\hat{H}_{A_{2}}$ belongs to the
one-dimensional irrep $A_{2}$ of the octahedron group $O$, while
the terms $\hat{H}_{F_{1}(i)}$ and $\hat{H}_{F_{2}(i)}$ ($i=1$,
2, 3) belong to its three-dimensional irreps $F_{1}$ and $F_{2}$
respectively, with ${a}_{2}$ and ${f}_{r\, i}$ ($r=1,2$;
$i=1,2,3$) being the Hamiltonian parameters.

After taking into account the simultaneous presence of quadrupole
degrees of freedom as well as the high order quadrupole--octupole
interaction we applied the above Hamiltonian to obtain the
rotational spectrum of the system by minimizing subsequently its
energy with respect to the third angular momentum projection $K$
for each given value of the angular momentum $I$.

We found that the so obtained energy bands after being used in
Eq.~(\ref{stag}) exhibit various staggering patterns in dependence
on the model parameters. Several schematic examples are
demonstrated in Fig.~1(b)-(d). On this basis we suppose that the
model can be applied to reproduce the staggering effects in
nuclear octupole bands \cite{DBoct00} as well as in some
rotational negative parity bands built on octupole vibrations .

\section{Conclusion}

The approaches suggested give a rather general prescription for
analysis of various fine characteristics of rotational motion in
quantum mechanical systems. They allow a detailed comparison of
the effects of the collective interactions and shapes in
different regions of nuclei. We propose a systematic study of the
symmetries associated with these effects and suggest that it
could provide a relevant guide in the revealing of some general
properties of collective motion in the less studied regions, such
as the exotic nuclei.

\section*{Acknowledgments}

This work has been supported by BNSF contract no MU--F--02/98.



\bigskip\bigskip

\centerline{\bf Figure captions}

{\bf Fig. 1} $\Delta I =1$
staggering patterns: (a)for the $\gamma$ band of $^{166}$Er
(theory and experiment); (b)--(d) for the schematic spectra
obtained by the octahedron Hamiltonian
[Eqs~(\protect\ref{HA})--(\protect\ref{HF23}), theory].

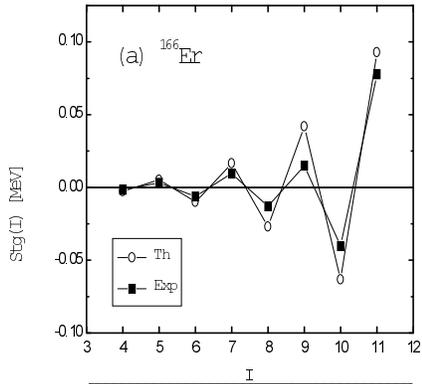
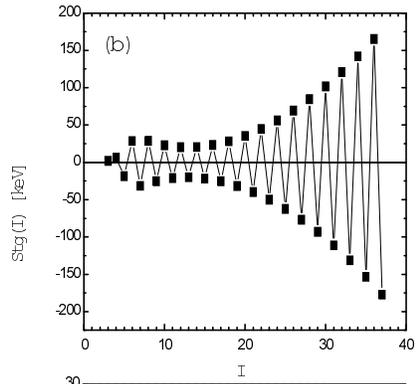
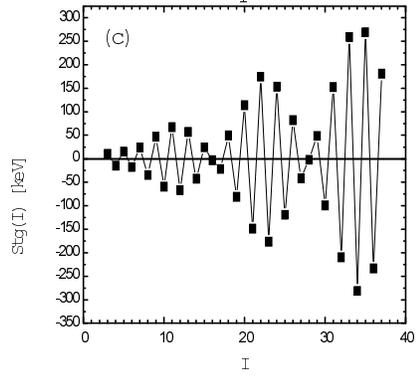
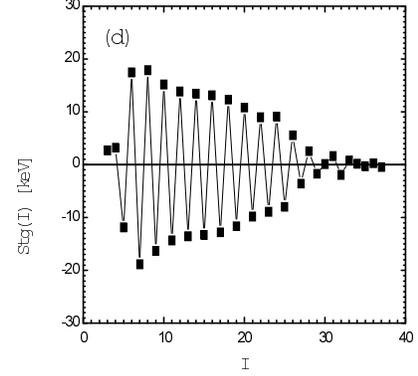
\end{document}